\documentclass[twocolumn,trackchanges]{aastex62}
\usepackage{amssymb}
\usepackage{graphicx}
\usepackage{grffile}
\usepackage{epsfig}
\usepackage{epstopdf}
\usepackage{url}
\usepackage{CJK}

\newcommand{\msun}{\ensuremath{\rm M_\odot}}

\newcommand{\secpoint}{\mbox{$''\mskip-7.6mu.\,$}}

\newcommand{\Ha}{\ensuremath{\rm H\alpha}}
\newcommand{\Hb}{\ensuremath{\rm H\beta}}

\newcommand{\lya}{\ensuremath{\rm Ly\alpha}}
\newcommand{\wlya}{$W_{\rm Ly\alpha}$}

\newcommand{\NHI}{$N_{\rm HI}$}

\newcommand{\kms}{km\,s\ensuremath{^{-1}}}

\newcommand{\HI}{\ion{H}{1}}

\newcommand{\fluxunits}{erg s$^{-1}$ cm$^{-2}$}

\newcommand{\sbunits}{erg s$^{-1}$ cm$^{-2}$ arcsec$^{-2}$}

\turnoffediting

\shorttitle{The Kinematics of Extended \lya\ Emission at $z=2.3$}
\shortauthors{Erb, Steidel \& Chen}

\submitjournal{ApJL}
\accepted{June 28, 2018}

\begin{document}
\begin{CJK*}{UTF8}{gbsn}

\title{The Kinematics of Extended \lya\ Emission in a Low Mass, Low Metallicity Galaxy at $z=2.3$\footnote{The data presented herein were obtained at the W. M. Keck Observatory, which is operated as a scientific partnership among the California Institute of Technology, the University of California and the National Aeronautics and Space Administration. The Observatory was made possible by the generous financial support of the W. M. Keck Foundation.}}

\correspondingauthor{Dawn K. Erb}
\email{erbd@uwm.edu}

\author[0000-0001-9714-2758]{Dawn K. Erb}
\affil{The Leonard E.\ Parker Center for Gravitation, Cosmology and Astrophysics, Department of Physics, University of Wisconsin-Milwaukee, 3135 N Maryland Avenue, Milwaukee, WI 53211, USA}

\author[0000-0002-4834-7260]{Charles C. Steidel}
\affiliation{Cahill Center for Astrophysics, California Institute of Technology, MC 249-17, 1200 E California Boulevard, Pasadena, CA 91125, USA}

\author[0000-0003-4520-5395]{Yuguang Chen}
\affil{Cahill Center for Astrophysics, California Institute of Technology, MC 249-17, 1200 E California Boulevard, Pasadena, CA 91125, USA}

\begin{abstract}

\lya\ photons are resonantly scattered by neutral hydrogen, and may therefore trace both the spatial extent and the kinematics of the gas surrounding galaxies. We present new observations of the extended \lya\ halo of Q2343-BX418, a low mass ($M_{\star} = 5\times10^8$ \msun), low metallicity ($Z\approx 0.25$ Z$_{\odot}$) star-forming galaxy at $z = 2.3$. Using the Keck Cosmic Web Imager, the blue-sensitive optical integral field spectrograph recently installed on the Keck II telescope, we detect \lya\ in emission to a radius of 23 kpc, and measure an exponential scale length of 6 kpc in the outer region of the extended halo. We study the double-peaked spectroscopic \lya\ profile in individual spectral pixels (``spaxels") over a $\sim 25\times30$ kpc region, finding significant variations in the peak ratio and peak separation. The profile is dominated by the red peak in the central regions, while in the outskirts of the extended halo the red and blue peak strengths are roughly equal; these observations are consistent with a model in which the peak ratio is largely determined by the radial component of the outflow velocity. We find a gradient of $300$ \kms\ in the \lya\ peak separation across the extended halo, indicating variations in the column density, covering fraction or velocity range of the gas. These new observations emphasize the need for realistic, spatially resolved models of \lya\ radiative transfer in the halos of galaxies.

\end{abstract}

\keywords{cosmology: observations---galaxies: evolution---galaxies: high-redshift}

\section{Introduction}
\label{sec:intro}

The luminous portions of galaxies are surrounded by an extended gaseous halo, the site of complex interplay between galactic outflows, inflows, and reaccretion. Evidence of this circumgalactic medium (CGM) is seen in enhancements of neutral hydrogen and metals near galaxies, detected in absorption in the spectra of background sources (e.g.\ \citealt{ass+05,ssrb06,ses+10,cbs+11,rst+12,tss+14,wpt+14}).

Further clues to the nature of the CGM may be found in the diffuse halos of \lya\ emission surrounding star-forming galaxies both locally \citep{hosv+13} and at high redshifts (\citealt{sbs+11,myh+12,mon+14,mon+16,wbb+16,lbw+17}; cf.\ \citealt{bfm+10,fhc+13}). The origin of these halos is debated, with suggestions including cooling radiation from cold accretion, emission from satellite galaxies, or resonant scatting of \lya\ photons from star formation by the CGM (e.g.\ \citealt{ts00,hsq00,fkd+10,zcw+11,lzc+15}). Recent observational studies have been unable to distinguish between these scenarios  \citep{lbw+17}.

Resonant scattering modifies the spectral profile of \lya\ emission as well as its spatial distribution. Spatially integrated \lya\ emission is nearly always redshifted relative to the systemic velocity, indicating the presence of a galactic outflow:\ \lya\ photons encounter the lowest optical depth when they backscatter from the receding outflow on the far side of the galaxy and thereby acquire a frequency shift enabling them to pass through the galaxy without further scattering. Galaxies with a relatively low \lya\ optical depth typically exhibit double-peaked profiles, with a dominant red peak and weaker blue peak bracketing the systemic velocity; the peak separation decreases with decreasing opacity, and the strength of the blue peak may increase as well (e.g. \citealt{vsm06,hsme15,gdmo16,vos+17}).

\lya\ emission is thus a powerful probe of the CGM, tracing both the spatial extent and the velocity structure of the gas surrounding galaxies. In practice, spatially resolved spectroscopic studies of the extended emission are difficult due to its low surface brightness; most work has examined the kinematics of the bright, extended \lya\ nebulae typically associated with multiple sources \citep{wbg+10,mcm+14,pmd15,svs+15,vbg+17} or the \lya\ emission of gravitationally lensed galaxies \citep{sbs+07,prv+16,ssm+17}. 

In this Letter we report new observations of the extended \lya\ emission of a low metallicity galaxy at $z=2.3$ with the Keck Cosmic Web Imager (KCWI; \citealt{kcwi1,kcwi2}), the blue-sensitive optical integral field spectrograph commissioned on the Keck II telescope in September 2017. Our target is Q2343-BX418 ($\alpha=23$:46:18.571, $\delta=+12$:47:47.379; J2000), a low mass ($M_{\star} = 5\times10^8$ \msun), low metallicity ($12+\log (\mbox{O/H})=8.08$) star-forming galaxy at $z_{\rm sys} = 2.3054$ with a high specific star formation rate ($\sim20$ Gyr$^{-1}$) and strong \lya\ emission. We target BX418 for KCWI observations because of its strong \lya\ emission, its potential similarity to galaxies in the reionization era for which the CGM properties may be of particular interest, and its wealth of existing data. BX418 has previously been examined in detail by \citet{eps+10}, with additional observations presented by  \citet{lse+09}, \citet{srs+14} and \citet{eps+16}. 

We describe the KCWI observations and data reduction in Section \ref{sec:data}, report our results in Section \ref{sec:results}, and discuss their implications in Section \ref{sec:discuss}.  We assume the  \citet{planck15} values of the cosmological parameters, $H_0=67.7$ \kms\ Mpc$^{-1}$, $\Omega_{\rm m}=0.31$, and $\Omega_{\Lambda}=0.69$; with these values, 1\arcsec\ subtends a distance of 8.4 kpc at $z=2.3$, the redshift of Q2343-BX418.

\begin{figure}[htbp]
\centerline{\epsfig{angle=00,width=\hsize,file=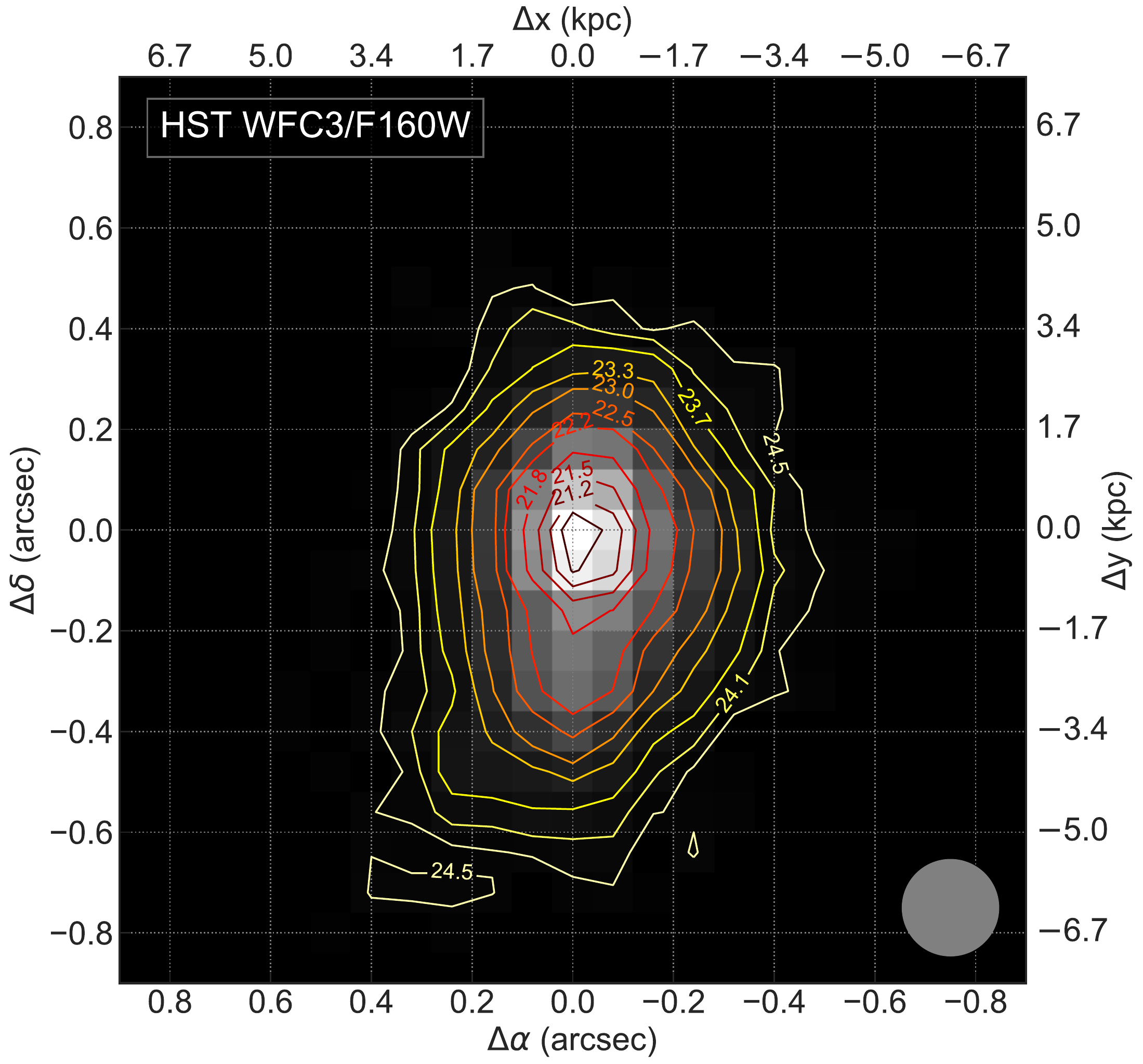}}
\caption{\textit{Hubble Space Telescope} WFC3/F160W image of Q2343-BX418. The rest-frame optical morphology is compact, with half-light radius 1.5 kpc \citep{eps+10}. Lower surface brightness emission is indicated by the contours, which mark surface brightness levels in AB magnitudes per square arcsecond ranging from 21.0 to 24.5. The 0.19\arcsec\ PSF is indicated by the grey circle at lower right.} 
\label{fig:hst_image}
\end{figure}

\begin{figure*}[htbp]
\centerline{\epsfig{angle=00,width=\hsize,file=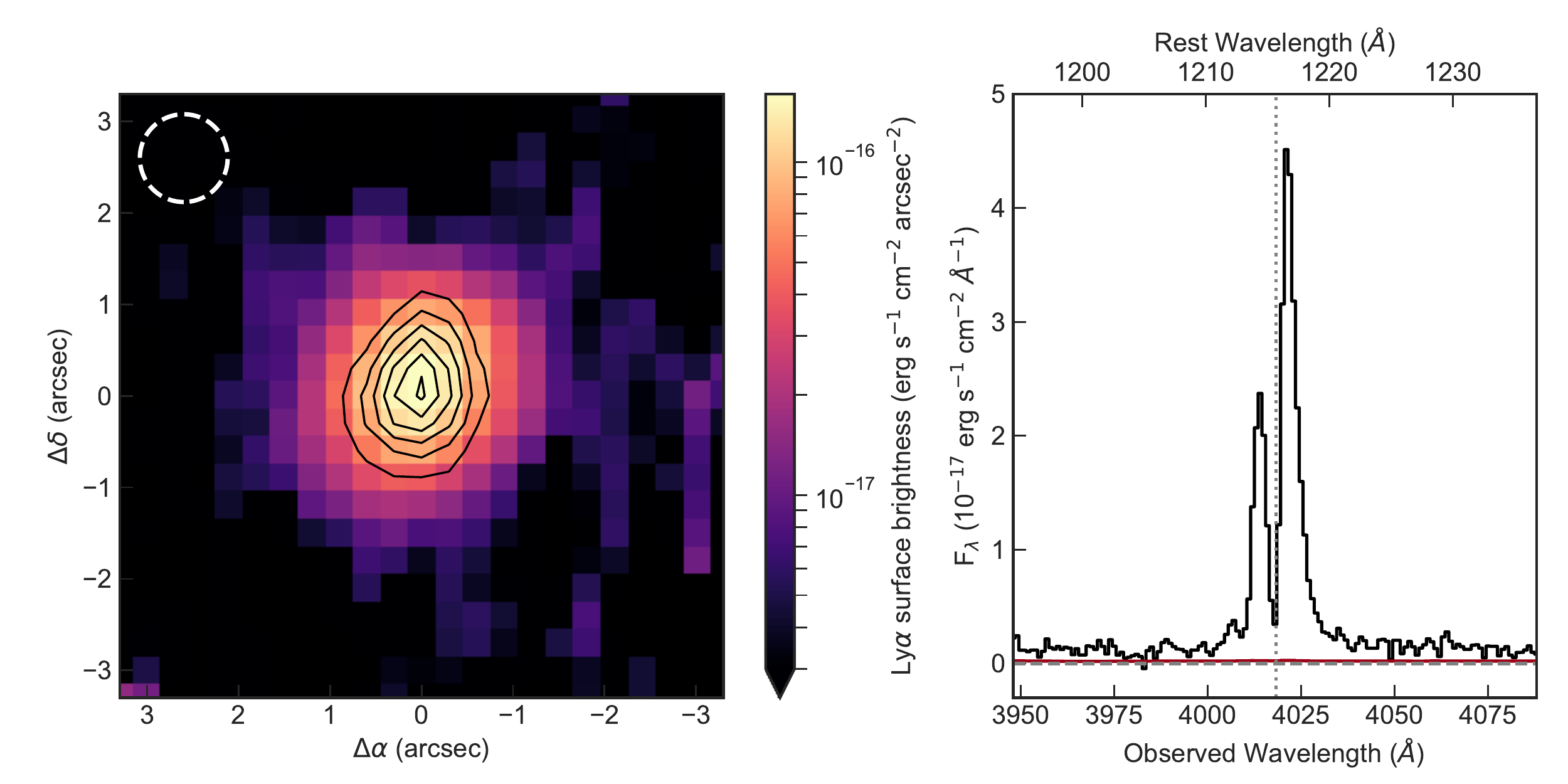}}
\caption{{\it Left:} Surface brightness of the continuum-subtracted \lya\ emission, shown on a logarithmic scale to emphasize faint emission at large radii. The dashed white circle in the upper left indicates the PSF with 1\arcsec\ FWHM, and contours indicate the surface brightness of the continuum image constructed as described in the text, with the lowest (highest) level at 0.6 (3.6)~$\times10^{-17}$ \sbunits.  {\it Right:} One-dimensional spectrum of the \lya\ emission extracted from the spatial sum of the central $8\times8$ spaxels ($2.4\arcsec\times2.4$\arcsec). The red line shows the 1$\sigma$ error spectrum.} 
\label{fig:lya_image}
\end{figure*}

\section{Observations and Data Reduction}
\label{sec:data}
Observations of Q2343-BX418 were obtained on the nights of 22--23 September 2017, under clear conditions with $\sim0.9$\arcsec\ seeing. 
The KCWI medium-scale slicer samples a contiguous field of 16\secpoint5$\times$20\secpoint4, where the longer dimension is along slices, with $24\times$0\secpoint69 samples perpendicular to the slices. The E2V 4k$\times$4k detector was binned 2$\times$2 on readout, resulting in spatial sampling on the detector of 0.29\arcsec\ along slices and 2.5 pixels per spectral resolution element. The KCWI-B BL grating was used with a camera angle placing 4500 \AA\ at the center of the detector; the common wavelength range sampled by all 24 slices is $3530-5530$ \AA, with a spectral resolution of 2.5~\AA\ (FWHM), for a resolving power of $R \sim 1400-2200$ depending on wavelength. 

Nine exposures of 1200~s each (total 10,800~s) were obtained with small telescope offsets between each to better-sample the spatial PSF in the direction perpendicular to slices. The KCWI Data Reduction Pipeline\footnote{\url{https://github.com/kcwidev/kderp}} was used to reduce raw CCD frames to wavelength calibrated, spatially rectified, differential atmospheric dispersion-corrected, background subtracted, flux calibrated data cubes with initial sampling of 0.29\arcsec\ by 0.69\arcsec\ (spatial) and 1 \AA\ pix$^{-1}$ (spectral). The individual reduced data cubes were then averaged with inverse variance weighting, after registration and spatial resampling onto an astrometrically-correct rectilinear grid sampled with 0.3\arcsec\ in each spatial dimension, rotated to place north up and east left. 

\section{The Extended \lya\ Halo of Q2343-BX418}
\label{sec:results}

We begin with the global \lya\ properties of Q2343-BX418. A one-dimensional spectrum extracted from the spatial sum of the central $8\times8$ spectral pixels reveals double-peaked \lya\ emission over the observed wavelength range 4002--4034 \AA\ (1210--1220 \AA\ in the rest frame; see Figure \ref{fig:lya_image}). We therefore collapse the data cube spectrally over this range to construct a \lya\ image. We also create a continuum image by collapsing a region of the same spectral width redward of the \lya\ line (4195--4227 \AA\ or rest-frame 1269--1279 \AA) and scaling it to the average value of the continuum measured on the blue and red sides of the line. We subtract this continuum image from the \lya\ image to create a line-only map of the \lya\ emission, shown in the left panel of Figure \ref{fig:lya_image}.

The continuum-subtracted \lya\ image reveals an extended halo of emission with peak surface brightness $\Sigma_{\rm max} = (1.71 \pm 0.026) \times10^{-16}$ \sbunits, and \lya\ is detected with S/N~$\geq3$ (1) per pixel to a radius of $2.0$ (2.4) arcsec or 16 (20) kpc.  The sum of the \lya\ image gives a total line flux $F_{\rm Ly\alpha} = (4.50 \pm 0.048) \times10^{-16}$ \fluxunits, corresponding to a luminosity $L_{\rm Ly\alpha} = (1.93\pm0.021) \times 10^{43}$ erg s$^{-1}$ at $z=2.3$. The total line flux divided by the average continuum level results in the rest-frame equivalent width \wlya~$=105 \pm 13$ \AA.  

We estimate the \lya\ escape fraction $f_{\rm esc}$ by comparing the \Ha\ and \lya\ fluxes, assuming an intrinsic ratio \lya/\Ha~$=8.7$. Recent measurements from Keck MOSFIRE indicate $F_{\rm H\alpha} = 1.3 \times 10^{-16}$ \fluxunits\ and \Ha/\Hb~$=3.3$ \citep{tss+18}; after correcting \Ha\ for extinction using the \citet{ccm89} extinction law, we find $f_{\rm esc}=0.28.$ However, this measurement is subject to significant systematic uncertainties related to the separate slit-loss corrections applied to \Ha\ and \Hb\ and the resulting uncertainties in the extinction correction. The \Ha\ flux measured from Keck OSIRIS, which is not subject to slit losses, is $8.2\times10^{-17}$ \fluxunits\ \citep{lse+09}; this value results in an escape fraction $f_{\rm esc}=0.45$--0.64, where the range brackets the uncertainties in extinction ranging from \Ha/\Hb~$=3.3$ as measured by MOSFIRE to zero extinction as implied by modeling of the spectral energy distribution. All estimates of the escape fraction are considerably less than unity, indicating that star formation within the galaxy is sufficient to produce the observed \lya\ emission.

\begin{figure*}[htbp]
\centerline{\epsfig{angle=00,width=\hsize,file=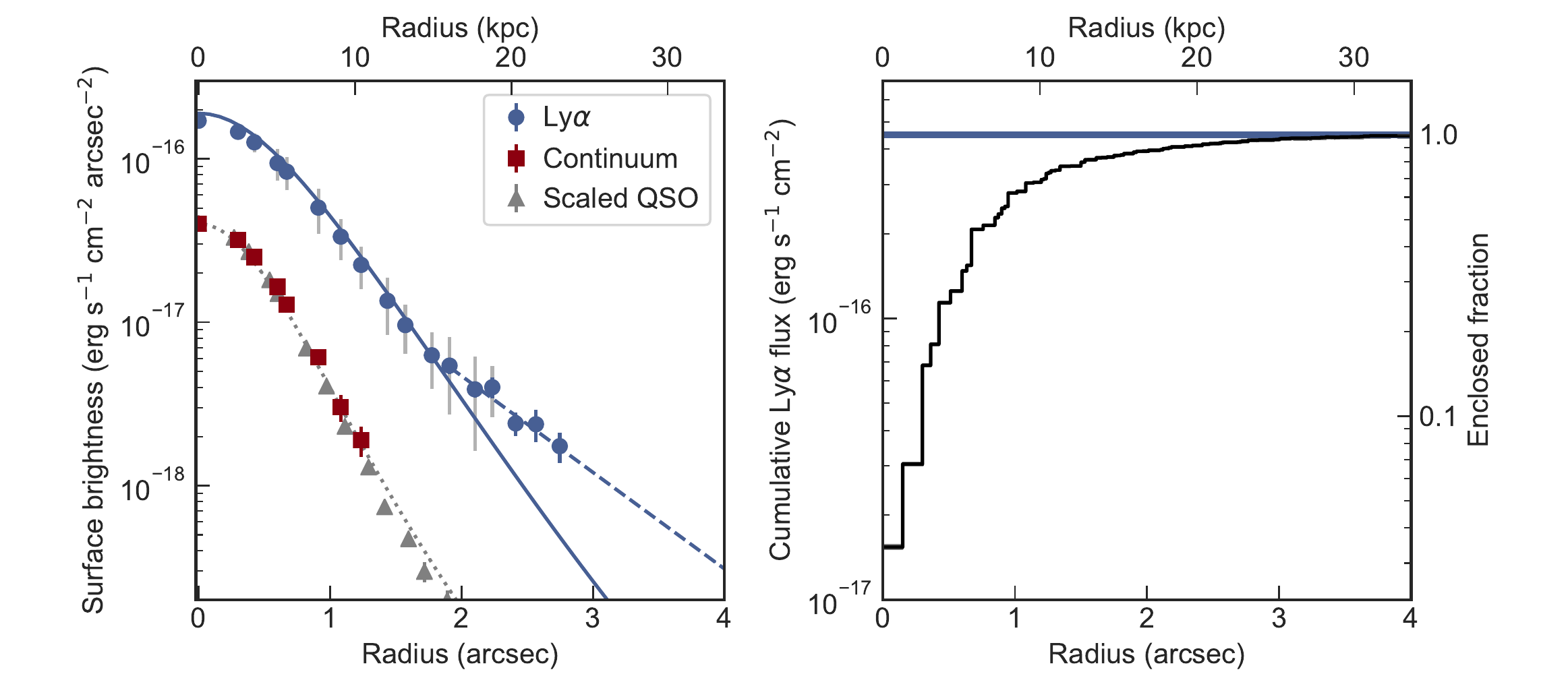}}
\caption{{\it Left:} Radial profiles of the \lya\ (blue) and continuum (red) emission. Error bars with colors matching the points indicate the error in the weighted mean, and larger grey error bars indicate the dispersion of individual spaxels in the \lya\ image. Only points with $>3\sigma$ detections are shown. Grey triangles show the radial profile of a QSO observed on the same nights as BX418, and the dotted grey line shows the best-fit Moffat profile to the QSO data. Blue lines indicate exponential fits as described in the text. {\it Right:} Growth curve for the total \lya\ flux as a function of radius. The horizontal blue line shows the total \lya\ flux measured as described in the text.}
\label{fig:radialprofiles}
\end{figure*}

\subsection{The Spatial Distribution of Ly$\alpha$ Emission}
\label{sec:spatial}
We quantify the spatial distribution of the light by constructing azimuthally averaged surface brightness profiles of both the \lya\ and continuum emission, radially binning the flux in 0.15\arcsec\ intervals. These profiles are shown in the left panel of Figure \ref{fig:radialprofiles}, where large symbols show the weighted mean in each bin and error bars of the same color show the error on the weighted mean. Larger grey error bars on the \lya\ profile indicate the dispersion of individual pixels. After binning, \lya\ emission is detected at $3\sigma$ to a radius of 2.7\arcsec\ or 23 kpc. 

\begin{figure*}[htbp]
\centerline{\epsfig{angle=00,width=\hsize,file=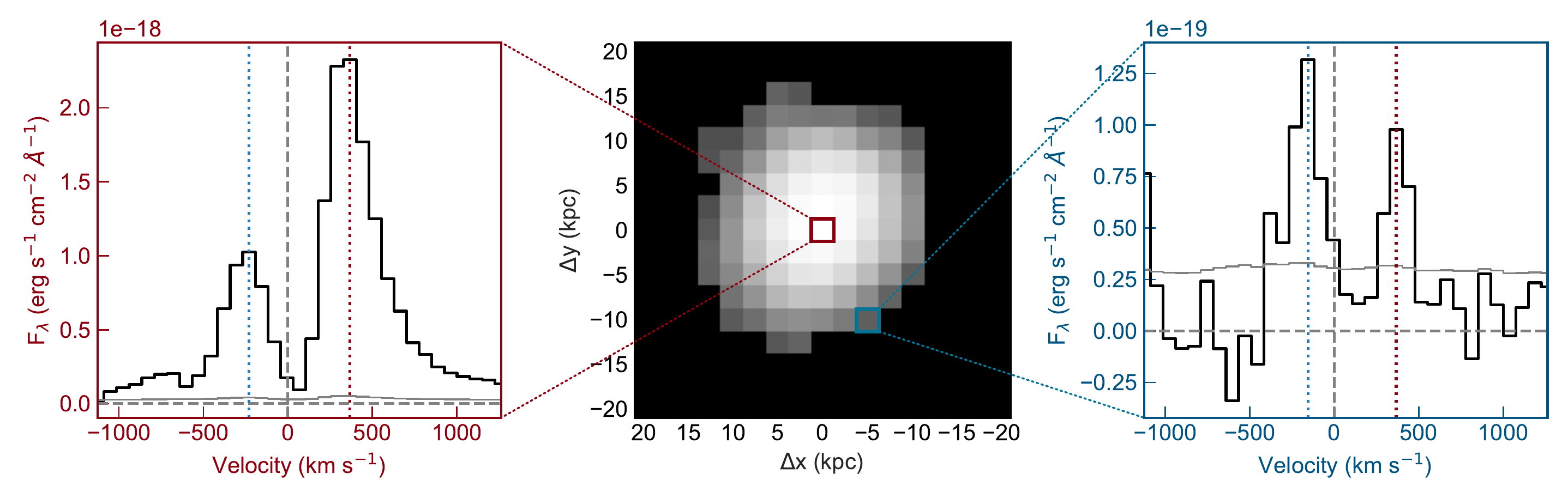}}
\caption{One-dimensional \lya\ velocity profiles from two different positions in the extended \lya\ halo. The central panel shows the logarithmic continuum-subtracted \lya\ surface brightness of all pixels with S/N~$\geq3.5$ (see Figure \ref{fig:lya_image} for numerical values), and spaxels at the center and edge of the halo are marked with red and blue boxes respectively. The left panel shows the \lya\ profile of the central spaxel, and the right panel shows the profile of the spaxel on the edge of the halo. In both panels, the solid grey line shows the 1$\sigma$ error spectrum and dotted blue and red vertical lines mark the positions of the blue and red peaks respectively.} 
\label{fig:aperturespectra}
\end{figure*}

\begin{figure*}[htbp]
\centerline{\epsfig{angle=00,width=\hsize,file=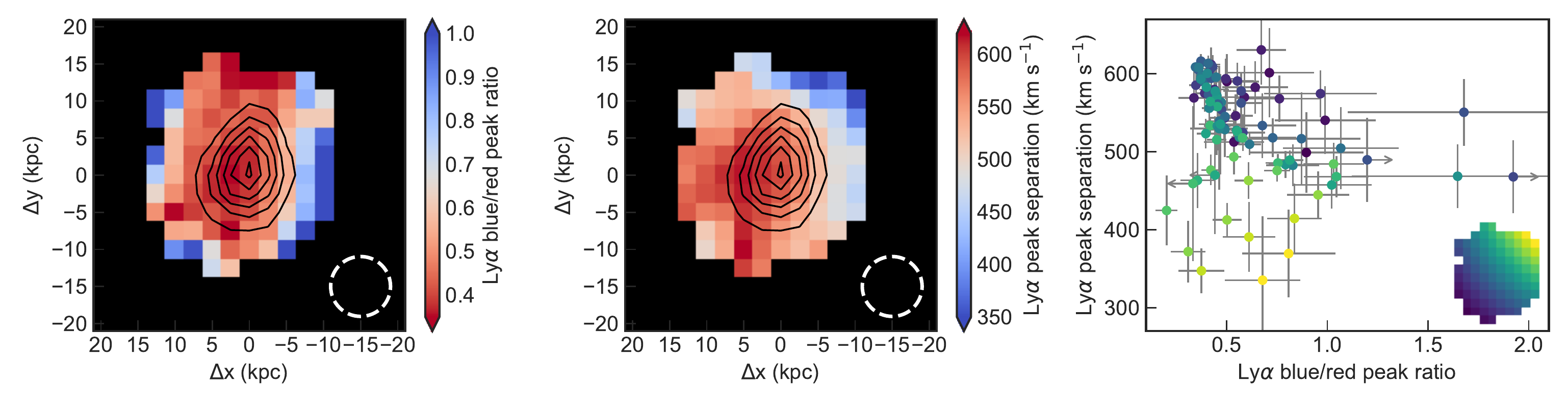}}
\caption{{\it Left:} Continuum-subtracted \lya\ emission color-coded by the blue/red flux ratio of the double peak. {\it Center}: Map of \lya\ peak separation $\Delta  v_{\rm peak} = v_{\rm red} - v_{\rm blue}$ in the extended halo. {\it Right:} \lya\ peak ratio vs.\ separation, with points color-coded by position in the halo as indicated by the inset map at lower right.}
\label{fig:lyamaps}
\end{figure*}

We also show the azimuthally-averaged profile of a quasar that was observed on the same nights as BX418, with exposures alternating with those of BX418 (grey triangles, where the peak value has been scaled to match the peak continuum surface brightness of BX418). The radial profile of the QSO is reasonably well-fit with a Moffat profile with FWHM~$=0.96$\arcsec\ (dotted grey line), and we use this profile as a model of the PSF.  As this model is a nearly perfect match to the continuum profile of BX418, we conclude that the continuum emission is not resolved; given the half-light radius of 1.5 kpc from HST imaging \citep{eps+10}, this is not surprising.

We then fit the binned \lya\ surface profile with an exponential function convolved with the PSF (solid blue line), finding a best-fit scale length of $2.58 \pm 0.02$ kpc; however, this model fits only the inner portion of the extended halo, significantly underestimating the \lya\ surface brightness at radii greater than 1.8\arcsec\ (15 kpc). Fitting an exponential function to only the points beyond this radius (dashed blue line), we find a scale length of $6.2 \pm 1.1$ kpc.

This scale length of 6.2 kpc lies at the low end of the scale lengths found for stacked \lya\ images at $z>2$, which range from $\sim7$ to $\sim25$ kpc \citep{sbs+11,mon+14,mon+16}; these measurements are also made at radii $\gtrsim2$\arcsec. It is also within the range of halo scale lengths found for individual galaxies detected by MUSE \citep{lbw+17}. 

In the right panel of Figure \ref{fig:radialprofiles} we show the growth curve for the \lya\ flux as a function of radius, with the total flux measured as described above indicated by the horizontal blue line. This curve indicates that half of the observed \lya\ emission lies within a radius of 7.1 kpc (0.8\arcsec), and the flux reaches 95\% of its total value at a radius of 23 kpc (2.8\arcsec).

\subsection{The Velocity Structure of Extended Ly$\alpha$ Emission}
\label{sec:spectral}
Examination of the \lya\ profiles of individual spaxels within the extended halo reveals significant differences in the blue-to-red peak ratio and velocity separation in different regions. We show two examples in Figure~\ref{fig:aperturespectra}. In the left and right panels of this figure we show one-dimensional \lya\ velocity profiles extracted from two different spaxels of the data cube. While both profiles are double-peaked, the central \lya\ emission is dominated by the red peak and has peak separation $\Delta v_{\rm peak} = 600$ \kms, while the emission from the edge of the halo has a blue-to-red peak ratio $\geq1$ and a somewhat narrower peak separation $\Delta v_{\rm peak} = 520$ \kms.

We find that we can make reliable measurements of the \lya\ profile for individual spaxels with S/N~$\geq3.5$ in the continuum-subtracted \lya\ image; for the wavelength ranges over which we spectroscopically detect \lya\ emission in these spaxels, we find \lya\ S/N~$>5$ for every spaxel with the exception of one with S/N~$=3.7$. The spaxels with S/N~$\geq3.5$ correspond to the $\sim25\times30$ kpc region shown in the center panel of Figure~\ref{fig:aperturespectra}. 

We measure the blue/red flux ratio by integrating the continuum-subtracted (for the spaxels in which the continuum is detected) \lya\ profile  of each spaxel between limits defined by the points where the data meet the error spectrum, separating the peaks at the wavelength of the trough between them. The spatial variation of this ratio is shown in the left panel of Figure \ref{fig:lyamaps}. As suggested by the profiles shown in Figure \ref{fig:aperturespectra}, the red peak dominates in the central region, while the peak ratio is roughly equal on the outskirts of much of the halo.

We also measure the separation between the two peaks by fitting a double Gaussian profile to the region between $-400$ and $+500$ \kms; this yields a determination of peak separation with uncertainty $<100$ \kms\ for all spaxels measured. The spatial distribution of the resulting peak separations $\Delta  v_{\rm peak} = v_{\rm red} - v_{\rm blue}$ is shown in the center panel of Figure \ref{fig:lyamaps}. We again observe large changes in the \lya\ profile as a function of position in the halo, with a range in peak separation of 300 \kms; the largest separations of $\sim500$--600 \kms\ occur in the southeast and central portions of the halo while the smallest separations of $\lesssim350$ \kms\ are found northwest of the central region. 

In the right panel of Figure \ref{fig:lyamaps} we plot the peak flux ratio vs.\ separation, color-coding the points by their position in the halo as shown by the small inset map. One-$\sigma$ error bars are shown for both quantities; in the few cases where the red or blue peak is formally undetected, we measure 3$\sigma$ lower and upper limits respectively. The spectrum shown in the right panel of Figure~\ref{fig:aperturespectra} corresponds to one of the outliers with a high peak ratio and large uncertainties.

The peak separation and flux ratio are significantly anti-correlated; we find a Spearman coefficient $-0.44$ and a probability of the null hypothesis in which the data are uncorrelated of $P=7.3\times10^{-6}\; (4.5\sigma)$. The color-coding of the points by position also shows that the correlation largely arises from the central regions of the halo, with outliers lying toward the southeast and northwest edges.

\section{Summary and Discussion}
\label{sec:discuss}
We have used the Keck Cosmic Web Imager to study the spatial and spectral properties of the extended \lya\ halo of Q2343-BX418 (Figure \ref{fig:hst_image}), a low mass ($M_{\star} = 5\times10^8$ \msun), low metallicity ($12+\log (\mbox{O/H})=8.08$) galaxy at $z=2.3$.  We detect \lya\ in emission to a radius of 23 kpc (see Figures \ref{fig:lya_image} and \ref{fig:radialprofiles}), and measure an exponential scale length of 6 kpc in the outer region of the extended halo. We measure the double-peaked \lya\ profile over a $\sim25\times30$ kpc region, finding that the flux ratio of the blue and red \lya\ peaks and their velocity separation vary significantly across the extended halo (Figures \ref{fig:aperturespectra} and \ref{fig:lyamaps}).

The spatial extent of the \lya\ emission from Q2343-BX418 is broadly consistent with many previous observations of star-forming galaxies at $z>2$ \citep{sbs+11,myh+12,mon+14,mon+16,wbb+16,lbw+17}, but little information regarding the velocity structure of such extended emission has so far been available. We therefore focus our discussion on the variations in \lya\ flux ratio and peak separation shown in Figure \ref{fig:lyamaps}.

The dominant red peak of most observed \lya\ profiles is a signature of galactic outflows, originating from \lya\ photons scattered by receding gas on the far side of the galaxy. The ubiquity of this profile in combination with the pervasive blueshifts of interstellar absorption lines suggests that outflows are poorly collimated at high redshifts, emerging from a typical galaxy in nearly all directions; this scenario is further supported by the lack of correlation between outflow velocity and apparent inclination \citep{lss+12b}.  \lya\ emission is a powerful diagnostic of this roughly spherical outflow because it does not rely on the backlight of the stellar continuum and can therefore reflect the velocity structure of the CGM gas at large radii. 

This simple outflow model provides a qualitative explanation for the \lya\ peak ratios observed in BX418. In the central regions of the \lya\ halo we expect to see a red-dominated \lya\ profile, as the bulk of the outflow velocity lies along our line of sight, while in the outer regions of the halo the outflow becomes largely transverse, resulting in blue and red peaks approximately equal in strength.  

The velocity separation between the blue and red peaks is widely interpreted as an indication of the \HI\ column density. Radiative transfer models indicate that the peak separation increases with increasing \NHI\ \citep{d14,vosh15}, and local galaxies with escaping Lyman continuum emission and therefore low \NHI\ are mostly observed to have double-peaked \lya\ emission with $\Delta v_{\rm peak} \lesssim 400$ \kms\ \citep{vos+17}. The increase in peak separation with increasing column density arises naturally if \lya\ photons in denser gas must be scattered to larger frequency shifts in order to escape. The \lya\ peak separation may also depend on the cloud covering fraction in a clumpy outflow, as a higher covering fraction may mimic the effects of an increase in column density \citep{gdmo16}.  This scenario implies a higher column density or covering fraction of outflowing gas in the central and southeast regions of BX418's \lya\ halo, where we observe peak separations $\Delta v_{\rm peak} \sim 500$--600 \kms. 

Alternatively, \citet{ses+10} develop a model for both \lya\ emission and the interstellar absorption lines, finding that the position of the \lya\ peaks is determined primarily by the velocity range of the gas; a wider velocity range results in both weaker emission and wider separation between the peaks. Thus \lya\ photons originating from regions of higher local velocity dispersion must scatter to larger velocities in order to escape their immediate environment. 
In the context of this model, it is striking to compare the variations in \lya\ peak separation with the AO-assisted mapping of the \Ha\ velocity dispersion presented by \citet[][see their Figure 2]{lse+09}. While the physical scales of the two measurements are very different, with the \Ha\ emission arising within $\sim 2 \times 3$ kpc, a gradient similar to that of the \lya\ peak separation is seen in the local \Ha\ velocity dispersion:\ the highest values of $\sigma \simeq 90$ \kms\ are seen in the southeast portion of the \Ha\ map, while the velocity dispersion decreases by nearly a factor of two to $\sigma \simeq 50$ \kms\ in the northwest. If the \lya\ peak separation reflects the velocity dispersion of the regions where the photons originate, 
this model then implies that some memory of this velocity distribution is preserved by photons re-scattered in regions of the outer halo.

A correlation between the local velocity dispersion and the \lya\ peak separation may also arise via connections between the velocity dispersion and the kinematics of outflowing gas; 
many numerical simulations of feedback in galaxies find that more efficient outflows are associated with higher gas velocity dispersions (e.g.\ \citealt{aklg13,adoo14}). Connections between \lya\ escape and the \Ha\ velocity dispersion are also seen in local galaxies; \citet{hgo+16} find that dispersion-dominated galaxies in the \lya\ Reference Sample (LARS) have higher \lya\ escape fractions than those with kinematics dominated by velocity shear.

These models depend on many factors and are not necessarily mutually exclusive; undoubtedly the column density, covering fraction and kinematics of neutral hydrogen all play roles in modulating the \lya\ emission in the extended halo. The observations of Q2343-BX418 we have presented here represent only one object, and while they are indicative of the complexities involved it remains to be seen whether the \lya\ kinematics presented here are typical. These data, and the larger sample of such observations still to come from the new optical integral field spectrographs, emphasize the need for realistic, spatially resolved models of \lya\ radiative transfer in the halos of galaxies.

\acknowledgements
The authors thank the referee for a thoughtful and constructive report, Max Gronke, David Kaplan, Aaron Smith, and Anne Verhamme for useful discussions and suggestions, KCWI team members Don Neill and Matt Matuszewski for their continued development of the KCWI DRP and for very useful conversations, and the organizers and participants of the ``Sakura Cosmic \lya\ Workshop" at the University of Tokyo in March 2018. DKE is supported by the US National Science Foundation through the Faculty Early Career Development (CAREER) Program, grant AST-1255591. CCS and YC acknowledge support from the Caltech/JPL President's and Director's Fund. The authors wish to recognize and acknowledge the very significant cultural role and reverence that the summit of Maunakea has always had within the indigenous Hawaiian community.  We are most fortunate to have the opportunity to conduct observations from this mountain.

\facility{Keck:II (KCWI)}
\software{KCWI DRP\footnote{\url{https://github.com/kcwidev/kderp}},
\texttt{astropy} \citep{astropy13}, 
\texttt{spectral{\textunderscore}cube} \citep{spectralcube}, 
\texttt{QFitsView\footnote{\url{http://www.mpe.mpg.de/~ott/QFitsView/index.html}}} ,
\texttt{seaborn} \citep{seaborn}}

\bibliographystyle{aasjournal}

\end{CJK*}
\end{document}